\documentclass[12pt]{article}
\usepackage{psfig}
\usepackage{bm}
\begin{document}
\def\He#1{{$\,{}^#1$He}}
\noindent{\LARGE\bf {Black hole and baby universe\\ in a thin film of
\He3-A}}\\
\addtocontents{toc}{ \protect\contentsline {section}
{{\sl Ted Jacobson and  Tatsuhiko Koike}}{}  } 
\bigskip

\noindent{{\bf Ted Jacobson }
\\[15pt]
University of Maryland\\
College Park\\
Maryland\\
USA\\
E-mail: jacobson@physics.umd.edu
\\[15pt]
{\bf Tatsuhiko Koike}
\\[15pt]
Keio University\\
Hiyoshi, Kohoku\\ 
Yokohama 223-8522\\
Japan\\
E-mail: koike@phys.keio.ac.jp
}
\vskip 1.0 cm
\noindent {\bf Abstract:} 
{\it Condensed matter black hole analogues may provide guidance in
grappling with difficult questions about the role of short distance
physics in the Hawking effect.  These questions bear on the very
existence of Hawking radiation, the correlations it may or may not
carry, the nature of black hole entropy, and the possible loss of
information when a black hole evaporates.  We describe a model of
black hole formation and evaporation and the loss of information to a
disconnected universe in a thin film of \He3-A, and we explain why the
existence of Hawking radiation has not yet been demonstrated in this
model.}

\vskip 1.5 cm
\noindent[Chapter for book, {\it Artificial Black Holes}, 
eds. M. Novello, M. Visser, and G. Volovik (World Scientific,
2002), based on a talk by TJ at the Workshop on Analog Models 
of General Relativity, held at CBPF in Rio de Janeiro, October
16-20, 2000.]
\def\3He{${}^3$He}
\def\ra{\rangle}
\def\la{\langle}
\def\ua{\uparrow}
\def\da{\downarrow}
\def\rar{\rightarrow}
\def\beq{\begin{equation}}
\def\eeq{\end{equation}}
\def\bea{\begin{eqnarray}}
\def\eea{\end{eqnarray}}
\def\O{\Omega}
\def\mi{i}
\def\scri{{\cal I}}
\def\eg{{\emph{e.g.}}}
\def\d{{\mathrm{d}}}
\section{Introduction and motivation}
From the condensed matter point of view, black hole analogues are a
curiosity. They certainly generate new questions, and they may lead to
new insights into the interplay between bulk and microscopic physics.
From the quantum gravity point of view, there is much more at
stake. We are looking to condensed matter systems for the guidance
they may provide in grappling with difficult questions about the role
of short distance physics in the Hawking effect. These questions bear
on the very existence of Hawking radiation, the correlations it may or
may not carry, the nature of black hole entropy\index{entropy}, and
the possible loss of information when a black hole evaporates.

Although the Hawking\index{Hawking effect} effect is a low-energy
phenomenon for black holes much larger than the Planck mass, it cannot
be deduced strictly within a low energy effective theory.  The
gravitational redshift from the event horizon is infinite, so the
outgoing modes that carry the Hawking radiation emerge from the
Planck\index{Planck regime} regime.  To derive the Hawking effect one
needs only the assumption that near the horizon these modes emerge in
their local ground state at scales much longer than the Planck length
but still much shorter than the black hole
radius~\cite{TJcutoff}. This assumption is {\it plausible}, since the
background is slowly varying in time and space compared to these
scales.  However, it is not {\it derived}.

A condensed matter analogy has already provided some support for this
picture.  Unruh~\cite{Ted-Unruh81} introduced a sonic\index{acoustic
black hole} analogue, in which the black hole is modeled by a fluid flow
with a supersonic ``horizon'', and the quantum field is replaced by
the quantized perturbations of the flow.  In a continuum treatment,
Unruh argued that the horizon would radiate thermal\index{thermal
phonons} phonons at the Hawking\index{Hawking temperature} temperature
$\hbar\kappa/2\pi$, where $\kappa$, which would be the surface gravity
for a black hole, is here the gradient of the velocity field evaluated
at the sonic horizon.  The short-distance (atomic) physics of this
analogue is fully understood in principle, hence it should be possible
to understand the origin and state of the Hawking\index{Hawking modes}
modes. As a first step inspired by this model, a number of studies
have been carried out where the linearity of the quantum field
equation is preserved but the short distance behavior is modified
either by introducing nonlinear dispersion\index{dispersion} or a
lattice\index{lattice} cutoff\index{cutoff}, designed to mimic some
aspects of the real atomic fluid.  The consequences have been
discussed in detail elsewhere (see~\eg~\cite{river} for a review).
The main point is that, despite the exotic origin of the outgoing
modes via ``mode\index{mode conversion} conversion'' near the horizon,
the short-distance physics does indeed deliver these modes in their
local ground state near the horizon if they originate far from the
horizon in their ground state.  These models thus lend some (linear
but nontrivial) support to the contention that Planck scale effects
deliver the local vacuum at a black hole horizon.

A controversial consequence of this simple picture, however, is that
the Hawking\index{Hawking effect} effect produces a loss of
information\index{information loss} from the world outside the
horizon.  The reason is that the local vacuum condition at the horizon
entails correlations between the field fluctuations inside and outside
the horizon.  The radiated Hawking\index{Hawking quanta} quanta are
thus correlated to ``partners'' that fall into the black
hole.\footnote{In addition to the information of correlations between
Hawking quanta and their partners, any information that simply falls
into the black hole from the outside appears to be lost.}  The origin
of these correlations is precisely the same as in the vacuum of flat
spacetime, so it is difficult to see any reason to doubt this
account. It is often doubted, however, since it implies that the
process of formation and complete evaporation of a black hole entails
non-unitary evolution in the Hilbert\index{Hilbert space} space
restricted to the outside world, which is considered by many (not
including the authors) to be a violation of quantum mechanics.  It
would be useful to have a down-to-earth condensed matter analogue in
which the information loss question arises but the fundamental physics
is understood.

Related to the issue of information loss is the nature of black hole
entropy\index{entropy}.  A spherical black hole of mass $M$ emitting
an energy $\d E=\d M c^2$ in thermal radiation at the
Hawking\index{Hawking temperature} temperature $T_H=\hbar c^3/8\pi GM$
loses an entropy $\d S = \d E/T_H = \d(A/4l_P^2)$, where $A=4\pi
R_s^2$ is the area of the event horizon of radius $R_s=2GM/c^2$, and
$l_P=(\hbar G/c^3)^{1/2}\simeq10^{-33} {\,\rm cm}$ is the Planck
length.  A black hole thus has one unit of entropy for every four
units of Planck area.  To understand the nature of the microscopic
degrees of freedom counted by this entropy remains one of the
outstanding problems of quantum black hole physics.  To solve this
problem will presumably require understanding the nature of the
short-distance cutoff\index{cutoff}.  Without a cutoff the entropy
would seem to be infinite due to the quantum
entanglement\index{entanglement} between field degrees of freedom on
either side of the event horizon discussed in the previous paragraph.
If a condensed matter horizon analogue produces Hawking\index{Hawking
effect} radiation, then it would seem to also carry an
entanglement\index{entanglement} entropy\index{entropy}, so it may
provide some guidance on the nature of black hole entropy.  (There are
important differences however, since in the condensed matter setting
there is no relation between the energy of the system and the area of
the horizon, the Einstein\index{Einstein equation} equation of course
does not pertain to the evolution of the horizon area, and the area
need not even change during the evolution of the system.)

The basic question of the existence of the Hawking\index{Hawking
effect} effect, as well as the issues of information\index{information
loss} loss and black hole entropy should be approachable in condensed
matter analogues.  The system we focus on here is particularly
interesting in that it provides a model of the formation and
evaporation of a black hole, with a disconnected part of the
``universe'', analogous to a so-called ``baby\index{baby universe}
universe'', into which information can potentially be lost.

It should be stated at the outset that at this stage we have only a
model of the background spacetime geometry.  A detailed analysis of
issues pertaining to the Hawking\index{Hawking effect} effect remains
to be carried out, and it is not yet clear that this system would
produce analogue Hawking\index{Hawking effect} radiation.  Another point
worth stressing is that, for the purposes of obtaining guidance in
quantum gravity issues, the experimental {\emph{feasibility}} of
actually setting up and observing a condensed matter analogue is not
essential.  Nevertheless, experimental observations are certainly one
of the goals of the whole program, both to confirm the basic
properties of the Hawking\index{Hawking effect} effect, and to gain
insight into the short-distance physics at play.

In the next section we introduce \He3-A and discuss how this medium
can be used to construct various black hole analogues.  This is followed
by a section focusing on the effective\index{effective geometry}
geometry and Hawking\index{Hawking effect} radiation in the moving
thin-film domain-wall\index{domain wall} model.  The last section
describes the analogue of black hole formation and evaporation and the
loss of information\index{information loss} to a disconnected universe
in the that model. We would like this article to be accessible to
researchers in both condensed matter and gravitational physics, hence
we include more than the usual amount of introductory material.  We
use units with $k_B=1$.

\section{Black hole analogues using \3He}

The Hawking\index{Hawking effect} effect is a quantum tunneling
process that produces a gentle instability of the ground state due to
the presence of unoccupied negative energy states in the
ergoregion\index{ergoregion} behind the horizon.  The instability
creates a flux of particles in a thermal state at the
Hawking\index{Hawking temperature} temperature.  If a condensed matter
system is to produce identifiable analogue Hawking\index{Hawking effect}
radiation, therefore, the system should presumably be at least be as
cold as the Hawking temperature,\footnote{This condition may be
avoided by observing instead a runaway quantum instability related to
the Hawking\index{Hawking effect} effect that is expected to occur for
bosonic fields when there is an inner horizon in addition to an outer
horizon~\cite{CJsuper,Anglin}.} which is very low temperature for
reasonable laboratory parameters.  Moreover, there should be no other
dissipation mechanisms that could swamp the Hawking\index{Hawking
effect} effect.

A natural place to start looking is therefore at
superfluid\index{superfluid} systems at zero temperature.  The case of
superfluid \He4\index{Helium~4} was initially examined in
\cite{ultrashort}, and further discussed in \cite{TJorigin}.  It was
concluded that a sonic horizon cannot be established in superflow,
because the flow is unstable to roton creation at the
Landau\index{Landau velocity} velocity which is some four times
smaller than the sound velocity.

\subsection{\3He-A}

Potentially more promising~\cite{JVhorizons,KVdecay,JVfilm,Vtorus} is
the (anisotropic) A-phase of superfluid\index{superfluid}
${}^3$He,\index{Helium~3-A} which has a rich spectrum of massless
quasiparticle\index{quasiparticle} excitations.  In particular, there
are fermionic quasiparticles\index{quasiparticle} --- the ``dressed''
helium atoms --- which have gapless\index{energy gap} excitations near
the gap nodes at $\vec{p}=\pm p_{F}\;\hat{\bm{l}}$ on the anisotropic
Fermi surface, and therefore can play the role of a massless
relativistic field\index{massless relativistic field} in a black hole
analogue.  The unit vector $\hat{\bm{l}}$ is the direction of orbital
angular momentum of the $p$-wave Cooper\index{Cooper pair} pairs and
$p_F$ is the Fermi\index{Fermi momentum} momentum.

For the benefit of readers not familiar with \3He-A we inject here a
lightning sketch of the basics.  (For complete introductions see
\cite{VollWolf,Vexotic,Leggett}.)  \3He is a spin-1/2 fermion, and is
described in a many-body fluid by a second quantized field operator
$\psi^A(x)$, where $A$ is a two-component spinor index.  The fluid has
a phase transition at 2.7 mK to a superfluid\index{superfluid} state
in which the order parameter $\la \psi^A({\bf x})\,\psi^B({\bf y})\ra
= f^{AB}({\bf x}-{\bf y})$ is non-zero.  Below a pressure of about 33
bars and a temperature of order 1 mK the fluid is in the so-called
B-phase.  The A-phase is a very long-lived metastable phase that is
coexistent with the B-phase and is stable in the region between 20 and
33 bars from 2.7 mK down to around 2 mK at the higher pressure.  In
the A-phase the order parameter, which can be thought of as the wave
function of a Cooper\index{Cooper pair} pair, is a spatial p-wave and
a spin triplet, and has the structure $|L=1, m_L=1\ra \otimes
|S=1,m_S=0\ra$.  The coherence\index{coherence length} length $\xi$,
which corresponds to the size of the Cooper\index{Cooper pair} pair
wavefunction, is of order 500 \AA.  The Fourier\index{Fourier
transform} transform of $f^{AB}({\bf x}-{\bf y})$ evaluated near the
Fermi\index{Fermi momentum} momentum $|{\bf p}|=p_F$ is proportional
to the energy gap\index{energy gap}
\beq 
\Delta({\bf p})={\Delta\over p_F}\; 
\left({\bf e}_1+i{\bf e}_2\right)^i\, p_i \;
{(|\ua\da\ra+|\da\ua\ra)\over\sqrt{2}},
\label{E:gap}
\eeq
where the three unit vectors ${\bf l}$, ${\bf e}_1$ and ${\bf e}_2$
form a right-handed orthonormal triad.  The gap\index{energy gap}
function (\ref{E:gap}) has nodes at $\vec{p}=\pm
p_F\;\hat{\bm{l}}$. Near these nodes the fermion
quasiparticles\index{quasiparticle} can have arbitrarily low energies
(above the Fermi\index{Fermi energy} energy), and they behave like
massless relativistic particles. The velocity of these quasiparticles
parallel to $\hat{\bm{l}}$ in ${}^3$He-A is the Fermi\index{Fermi
velocity} velocity $v_F\sim 55$ m/s, while their velocity
perpendicular to $\hat{\bm{l}}$ is only $c_\perp =\Delta/p_F\sim 3$
cm/s, where $\Delta\sim T_c\sim 1$ mK is the energy gap\index{energy
gap}.

\subsection{Black hole candidates}
\subsubsection{Superflow}
It should be possible to set up an inhomogeneous superflow exceeding
the slow speed $c_\perp$ in a direction normal to $\hat{\bm{l}}$, thus
creating a horizon for the fermion
quasiparticles\index{quasiparticle}.  There is a catch, however,
since the superflow is unstable when the speed relative to a container
exceeds $c_\perp$~\cite{KVdecay}.  A possible way around this was
suggested by Volovik~\cite{Vtorus}, who considered a thin film of
\He3-A flowing\index{flowing fluid} on a substrate of superfluid\index{superfluid} \He4, 
which insulates the \He3 from contact with the container.  In such a
film, the vector ${\bf l}$ is constrained to be perpendicular to the
film.  A radial flow on a torus, such that the flow velocity near the
inner radius exceeds $c_\perp$, would produce a horizon.
Theoretically this looks promising, however the Hawking\index{Hawking
temperature} temperature for a torus of size $R$ is $T=
(\hbar/2\pi)(\d v/\d r)\sim \hbar c_\perp/R\sim (\lambda_F/R)$ mK,
where $\lambda_F$ is the Fermi wavelength, which is of the order of
Angstroms.  Thus, even for a micron sized torus, the
Hawking\index{Hawking temperature} temperature would be only $\sim
10^{-7}$ K.

\subsubsection{Moving solitonic texture}
An alternative is to keep the superfluid\index{superfluid} at rest
with respect to the container, but arrange for a texture in the order
parameter to propagate in such a way as to create a horizon.  For
example, in~\cite{JVhorizons} a moving ``splay
soliton''\index{soliton} is considered.  This is a planar
texture\index{texture} in which the $\hat{\bm{l}}$ vector rotates from
$+\hat{x}$ to $-\hat{x}$ along the $x$-direction perpendicular to the
soliton\index{soliton} plane.  A quasiparticle\index{quasiparticle}
moving in the $x$-direction thus goes at speed $v_F$ far from the
soliton and at speed $c_\perp$ in the core\index{core} of the soliton.
If the soliton is moving at a speed greater than $c_\perp$, the
quasiparticles will not be able to keep up with it, so an effective
horizon will appear.  This example turns out to be rather interesting
and complicated in the effective relativistic description.  The null
rays on the horizon have a transverse velocity, making it like that of
a rotating black hole rather than a static black hole.  In addition,
since the $\hat{\bm{l}}$ vector couples to the
quasiparticles\index{quasiparticle} like an electromagnetic vector
potential, its time and space dependence generates a strong
``pseudo-electromagnetic'' field outside the ``black hole'' which
would produce quasiparticle\index{quasiparticle} pairs by analogy
with Schwinger pair production~\cite{Vexotic}. (This latter process
may be the same as what produces the so-called ``orbital
viscosity''~\cite{VollWolf} of a time-dependent
texture\index{texture}.)  The Hawking\index{Hawking temperature}
temperature also tends to be very low, and it seems likely that the
Hawking\index{Hawking effect} effect would be masked by the
pseudo-Schwinger pair production, though this has not been
definitively analyzed.

\subsubsection{Thin film with a moving domain wall}
In \cite{JVfilm} a simpler system was studied, that of a thin film of
\He3-A, perhaps on a \He4 substrate, with a domain\index{domain wall}
wall.  The vector $\hat{\bm{l}}$, which is perpendicular to the film,
has opposite sign on either side of the wall, and in the wall region
the condensate\index{condensate} is in a different
superfluid\index{superfluid} phase.  At the core\index{core} of the
wall the group velocity of the quasiparticles\index{quasiparticle}
goes to zero, so if the wall itself is moving, a horizon will appear.
For the rest of this article we focus on this example.  The effective
spacetime geometry of this system was first studied in~\cite{JVfilm}.
Here we extend that analysis to the case of a wall that accelerates
and then comes to rest again.

\section[Effective spacetime from a moving domain wall]
{Effective spacetime and Hawking effect from a moving domain wall texture}
\subsection{Texture and spectrum}

\begin{figure}[htb]
\bigskip
\centerline{
\psfig{figure=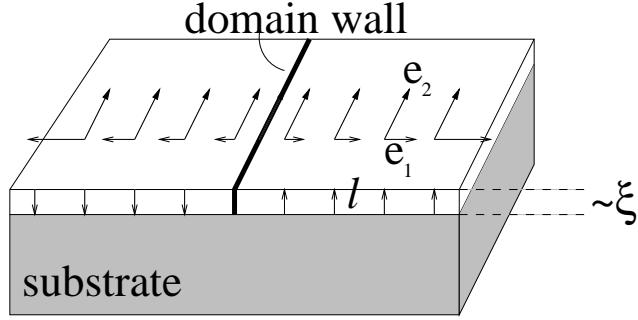,angle=-90,height=4cm}}
\caption[Thin-film domain-wall texture]{\sl Thin-film domain-wall texture.}
\label{F:film}
\end{figure}

The order parameter for a domain\index{domain wall} wall
texture\index{texture} in a thin film is described by a
gap\index{energy gap} function of the same form as (\ref{E:gap}), with
the unit vector ${\bf e}_1$ replaced by something like $\tilde{\bf
e}_1=\tanh(x/d)\;{\bf e}_1$, where $d\sim\xi$ is the width of the
wall.  Here it is assumed that the film lies in the x-y plane with the
wall along the y-axis.  (See figure~\ref{F:film}) The thickness of the
film should be not much more than the coherence\index{coherence
length} length $\xi$ in order for the domain\index{domain wall} wall
to be stable.

The quasiparticle\index{quasiparticle} spectrum in the A-phase takes
the form
\beq
E^2({\bf p})=v_F^2\;(p-p_F)^2+c_\perp^2\;({\bf e}_1\cdot{\bf p})^2+
c_\perp^2\;({\bf e}_2\cdot{\bf p})^2.
\label{E:spectrum} 
\eeq 
There is no excitation perpendicular to the thin film, hence we have
$p_z=p_F$, so
\beq 
p=\sqrt{p_F^2+p_x^2+p_y^2}=p_F+ \frac{1}{2p_F}(p_x^2+p_y^2)+\cdots.  
\eeq 
Using this expansion in (\ref{E:spectrum}) together with the
replacement ${\bf e}_1\rightarrow\tilde{\bf e}_1$ we obtain the low
energy spectrum for motion in the $x$--$y$ plane in the
domain\index{domain wall} wall texture\index{texture}:
\beq
E^2= c(x)^2\;p_x^2+c_\perp^2\;p_y^2 + 
\frac{c_\perp^2}{\hbar^2}\;\xi^2\;(p_x^2+p_y^2)^2+\cdots.
\label{E:xyspectrum}
\eeq 
In the quartic term we have replaced $v_F/2\Delta$ by $\xi/\hbar$, to
which it is roughly equal.  The speed $c(x)$, defined by
\beq c(x)=c_\perp \tanh(x/d),
\label{E:c(x)}
\eeq
goes to zero at the core\index{core} of the domain\index{domain wall} wall.

In the low energy limit the quartic term is negligible and the
spectrum becomes that of a massless relativistic particle in two
dimensions.  The nonrelativistic quartic corrections become important
at higher energy, when the wavelength is of order the
coherence\index{coherence length} length $\xi$.  Note that this is of
order 500 \AA, much longer than the interatomic spacing.\footnote{In
the analogy with quantum gravity, it would appear that the Planck
scale should be identified with $\xi$ since this measures the
``elasticity'' of the background, and there is at present no analogue
of the underlying atomic scale in fundamental theory except perhaps
the string scale, which is usually taken to be {\it longer} than the
Planck length, rather than shorter.} The corrections produce
``superluminal''\index{superluminal} group velocities at high
momentum.  If a quasiparticle\index{quasiparticle} is {\it localized}
near the domain\index{domain wall} wall then these nonrelativistic
corrections are important, since the width of the wall is of order
$\xi$.

\subsection{Spacetime of the stationary wall}
\label{S:statwall}
The relativistic limit of (\ref{E:xyspectrum}) is that of a massless
particle in a 2+1 dimensional spacetime with the line
element\index{line element}
\beq
\d s^2=-\d t^2 + c(x)^{-2}\d x^2 + c_\perp^{-2}\d y^2.
\eeq
The metric has translation invariance in the y-direction, so we will
make a ``dimensional reduction'' to the 1+1 dimensional spacetime
\beq
\d s^2 = -\d t^2 + c(x)^{-2}\d x^2.
\label{E:rest}
\eeq
Clearly this spacetime is {\it flat}, since one can introduce a new
spatial coordinate by $\d x_*=\d x/c(x)$ in terms of which the line
element takes the manifestly flat form $\d s^2 = -\d t^2 +\d
x_*^2$. Note however that since $c(x)$ goes to zero linearly as
$x\rightarrow0^+$, the coordinate $x_*$ goes to $-\infty$
logarithmically as the domain\index{domain wall} wall is approached
from the side of positive $x$. Therefore the film really corresponds
to {\it two complete copies} of flat spacetime, joined ``at infinity''
at the wall.

\subsection{Spacetime of the moving wall}
\label{S:movewall}
\begin{figure}[htb]
\bigskip
\centerline{
\psfig{figure=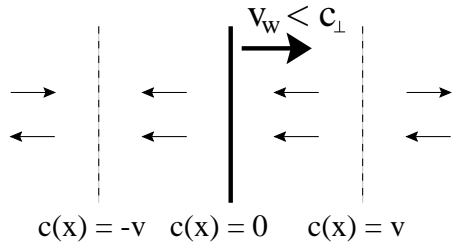,angle=0,height=3cm}}
\caption[Quasiparticles in a moving domain-wall texture]
{\sl Quasiparticles in a moving domain-wall texture.}
\label{F:vwall}
\end{figure}

Now suppose the domain\index{domain wall} wall texture\index{texture}
is moving to the right at speed $v<c_\perp$ relative to the
superfluid\index{superfluid} condensate\index{condensate}.  Then right
moving quasiparticles\index{quasiparticle} sufficiently far from the
wall will stay ahead of the wall, but those inside the point where
$c(x)=v$ will fail to stay ahead.  There will be a black hole horizon
where $c(x)=v$ and a white hole\index{white hole} horizon where
$c(x)=-v$ on the left hand side of the wall.  In between the two
horizons all low energy quasiparticles\index{quasiparticle} are
purely left-moving relative to the wall texture\index{texture} (see
figure~\ref{F:vwall}).  The closer $v$ is to $c_\perp$ the farther
apart the horizons lie.

To determine the spacetime metric for the moving wall, we introduce
coordinates $x_s$ and $x_w$ at rest with respect to the
superfluid\index{superfluid} and the wall, respectively.  These are
related by the Galilean\index{Galilean relativity} transformation
$x_s=x_w+vt$.  The dispersion\index{dispersion relation} relation is
determined in the superfluid\index{superfluid} frame, so the line
element\index{line element} (\ref{E:rest}) applies in the superfluid
frame, however the argument of the function $c(x)$ should be $x_w$
since this function describes the texture\index{texture} which is at
rest in the wall frame.  For the moving wall we thus have
\bea
\d s^2
&=&-\d t^2+c(x_w)^{-2}\d x_s^2\\
&=&-\d t^2+c(x_w)^{-2}(\d x_w+v\d t)^2\\
&=&-(1-v^2c(x_w)^{-2})\d t^2 + 
2vc(x_w)^{-2}\,\d t\,\d x_w +c(x_w)^{-2}\d x_w^2.
\label{E:movingwall}
\eea
Perhaps surprisingly, this is no longer a flat spacetime.  It has
black and white hole\index{white hole} horizons at $c(x_h)=\pm v$.
The wall core\index{core} at $x_w=0$ is now a {\it spacelike} line
(since the coefficient of $\d t^2$ is positive there), and it lies at
{\it finite} proper time along geodesics.  The Ricci curvature scalar
is given by
\beq
R=\frac{-4v^2}{d^2}\left(\frac{c_\perp^2}{c(x_x)^2}- 
\frac{c(x_x)^2}{c_\perp^2}\right).
\eeq
This diverges like $-(2v/x_w)^2$ as $x_w=0$ is approached, so there is
a curvature singularity at the core\index{core}.  The spacetime
therefore looks rather like that of an eternal Schwarzschild black
hole.  The curvature at the horizon is given by
\beq
R_{\mathrm{horizon}} = -(2c_\perp/d)^2\;[1-(v/c_\perp)^4].
\eeq

Unlike the maximally extended Schwarzschild black hole, however, the
spacetime of the moving wall is {\it incomplete}, in that geodesics
can run off the edge of the coordinate system $(t,x_w)$ in a finite
proper time or affine parameter.  The location of the incomplete
boundaries will be indicated below.  Of course physical
quasiparticles\index{quasiparticle} cannot escape, because this
really is the entire physical spacetime.  What happens is that as a
quasiparticle heads in the direction of an incomplete boundary (either
forward or backward in time), it is blueshifted into the part of the
spectrum where the nonrelativistic corrections become important, at
which point it propagates superluminally\index{superluminal} and the
geodesics of the effective metric no longer determine its trajectory.

\subsection{Hawking effect}
\label{SS:hawking}
Since the effective spacetime is that of a black hole, it is natural
to suppose the horizon would radiate fermion
quasiparticles\index{quasiparticle} at the Hawking\index{Hawking
temperature} temperature $T_H=(\hbar/2\pi)\kappa$, where $\kappa=\d
c/\d x(x_h)$ is the surface gravity of the horizon.  For the metric
(\ref{E:movingwall}) with (\ref{E:c(x)}) we have explicitly
\beq
T_H(v)=T_H(0)\;(1-v^2/c_\perp^2),\qquad\qquad  
T_H(0)={\hbar c_\perp\over2\pi d}.
\label{E:TH}
\eeq
Putting in the numbers we have $T_H(0)\sim 1 \;\mu$K. Equation
(\ref{E:TH}) gives the Hawking\index{Hawking temperature} temperature
in the wall reference frame (the ``Killing\index{Killing temperature}
temperature'' in the language of general relativity), which is related
to the temperature in the asymptotic frame of the
superfluid\index{superfluid} by a Doppler\index{Doppler effect} shift
factor:
\beq
T_{H, sf}=T_H(0)\;(1+v/c_\perp).
\label{E:Tsf}
\eeq
The Hawking\index{Hawking temperature} temperature in the frame of the
superfluid\index{superfluid} is thus never less than $T_H(0)$.
Although this is three orders of magnitude below the
critical\index{critical temperature} temperature, and extremely low in
practical terms, it is nevertheless close to where the
non-relativistic corrections become important (assuming $d\sim\xi$).

While the black hole analogy looks compelling, it should be emphasized
that the Hawking\index{Hawking effect} effect depends on behavior of
the quantum field that may not be valid in this context.  As discussed
in the introduction, the required condition is that the high frequency
outgoing modes near the horizon be in their quantum ground state.  In
this case these modes come from the singularity, since they propagate
``superluminally''\index{superluminal}.  The propagation of these
modes though the singularity may excite them.  This has not yet been
analyzed.\footnote{However the related problem of
quasiparticle\index{quasiparticle} tunneling across the stationary
domain\index{domain wall} wall has been studied by Volovik in section
11.1 of \cite{VolovikPR}.} If they are excited, this may suppress the
Hawking\index{Hawking effect} radiation by virtue of the
Pauli\index{Pauli principle} principle.  (Had the field been bosonic,
the excitations would have produced extra, induced emission.)  Another
difference from the black hole case is that once a
Hawking\index{Hawking pair} pair is produced, the negative energy
partner rattles around trapped in the ergoregion\index{ergoregion}
between the two horizons.  Once these available states fill up further
Hawking\index{Hawking effect} radiation would be suppressed.  (In the
case of a black hole, by contrast, the negative energy partners fall
into the singularity never to return.) For a discussion of aspects of
the behavior of superfluids\index{superfluid} in the presence of a
quasiparticle\index{quasiparticle} horizon see
reference~\cite{VolovikPR}.

For the remainder of this article we assume that, in spite of
Pauli-blocking effects, the moving domain\index{domain wall} wall
produces at least some Hawking\index{Hawking effect} radiation, and we
go on to study the analogue of the process of formation and
evaporation of a black hole. According to (\ref{E:TH}) the
Hawking\index{Hawking temperature} temperature approaches a nonzero
constant as $v\rightarrow0$. Nevertheless it is clear that at $v=0$
there can be no radiation since the wall is static and there is no
horizon, hence there is no ergoregion\index{ergoregion} with negative
energy states to be filled. The radiation rate must therefore go to
zero as $v$ goes to zero. If it goes as a power of $v$ then we have
$\d E/\d t = -b\;v^n$. The kinetic energy of the moving
domain\index{domain wall} wall is proportional to $v^2$ if, as seems
plausible, the action is dominated by quadratic terms in the time and
space derivatives. In this case $E=\mu v^2/2$ for some constant
$\mu$. Integrating the energy loss, we find that it takes a finite
time for the wall to come to rest if $n<2$, but an infinite time if
$n\ge2$.

Finally, a comment about entropy\index{entropy}.  It is tempting to
try and define a thermodynamic entropy\index{entropy} $S$ for the
moving domain\index{domain wall} wall, however it is by no means clear
that this should be meaningful.  In analogy to the black hole entropy,
one might define $S$ via $\d S = \d E/T_{H, sf}$, where $E= \mu v^2/2$
as above and $T_{H,sf}$ is the Hawking\index{Hawking temperature}
temperature in the superfluid\index{superfluid} frame, (\ref{E:Tsf}).
This yields the formula $S=(2\pi\mu c_{\perp}d/\hbar)\;
\left(v/c_\perp -\ln(1+v/c_\perp)\right)$.  This analogy seems flawed
however, as the domain\index{domain wall} wall is not stationary in
the superfluid\index{superfluid} frame so does not represent an
``equilibrium'' system.  If we try to correct this by passing to the
frame of the moving wall, we run into the problem that, as the wall
slows down due to Hawking\index{Hawking effect} radiation, this
comoving frame {\it changes}, unlike in the black hole case where the
asymptotic rest frame of the black hole is fixed even as the black
hole evaporates (or absorbs radiation).

\section[Hawking effect in the thin-film domain-wall model]
{Black hole formation and evaporation in the 
thin-film domain-wall model}
\subsection{Carter--Penrose causal diagrams}

To best exhibit the incompleteness of the black hole spacetime
discussed in the previous section, as well as features of the case
where the hole forms and then evaporates, it is helpful to use
Carter--Penrose diagrams.  We therefore pause here to explain what
such a diagram is for the benefit of readers from the condensed matter
side.

The basic idea is to draw a picture representing the causal structure
of a spacetime by showing light rays at 45$^\circ$, with regions at
infinite time or space brought into a finite location by a spacetime
dependent conformal rescaling of the line element\index{line element},
$\d\tilde{s}^2 =\O^2 \; \d s^2$, where $\O\rar 0$ at infinity.  Since
the causal structure is determined by the light cones, which are the
solutions of $\d s^2=0$, the causal structure of $\d\tilde{s}^2$ is
identical to that of $\d s^2$.  (See for example~\cite{HawkEllis}.)

\begin{figure}[tbh]
\centerline{
\psfig{figure=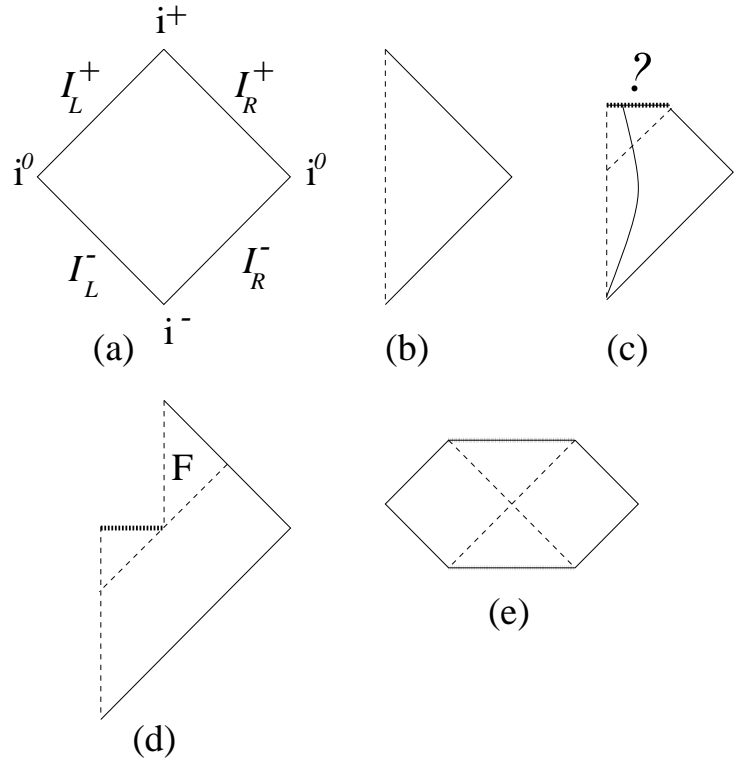,angle=0,height=10cm}}
\caption[Carter--Penrose diagrams]{\sl Carter--Penrose diagrams.}
\label{F:CP}
\end{figure}

As an example, consider 1+1 dimensional flat spacetime given by the
line element\index{line element} $\d s^2=-\d t^2 +\d x^2 = -\d u \, \d
v$, where $u=t-x$ and $v=t+x$.  The conformal factor $\O(u,v)=(\cosh u
\cosh v)^{-1}$ brings infinity into a finite location in the sense
that the points at infinity for $\d s^2$ lie at a finite proper distance
for $\d \tilde{s}^2$.  A diagram of the tilde spacetime is then just a
diamond, figure~\ref{F:CP}\,(a).  The boundaries of the spacetime are
at infinite time and/or space.  Timelike geodesics (straight lines in
this case) emerge from past timelike infinity $\mi^-$ and terminate at
future timelike infinity $\mi^+$.  Spacelike geodesics go from left
spacelike infinity $\mi^0_L$ to right spacelike infinity $\mi^0_R$,
and null geodesics or light rays go from left or right past null
infinity $\scri^-$ (``scri-minus'') to right or left future null
infinity $\scri^+$ (``scri-plus'').

In four-dimensional flat spacetime the spherical symmetry can be used
to reduce to a two dimensional diagram.  In spherical coordinates the
metric is $\d s^2=-\d t^2 + \d r^2 + r^2(\d\theta^2 + \sin^2\theta\,
\d\phi^2)$.  If we now define $u =t-r$ and $v=t+r$, the geometry of
the $t$-$r$ subspace for each set of polar angles is identical to the
1+1 dimensional case except that now only $v>u$ is physical.  The
spacetime is thus represented by figure~\ref{F:CP}\,(b), a diagram
that is half of a diamond, with each point representing a 2-sphere
except those on the vertical line on the left, which represents $r=0$.
This is shown as a dotted line.

The spacetime of spherical matter that collapses to form a black hole
looks like figure~\ref{F:CP}\,(c).  The shaded region represents the
collapsing matter.  The dashed line represents the event horizon, and
the thick-dashed line represents the curvature singularity inside the
black hole.  Note that the singularity is {\it spacelike}, and is the
future terminus of any causal curve that goes beyond the horizon.  It
is unknown how or even if spacetime develops in any form to the future
of the singularity, so a question mark is placed there.  One common
hypothesis is that a ``baby\index{baby universe} universe'' is born
there, which is disconnected from the outside world.  A controversial
question is whether such a baby\index{baby universe} universe can
harbor information\index{information loss} unavailable to the outside
world.

The diagram for a black hole that forms by collapse and then
evaporates away is shown in figure~\ref{F:CP}\,(d). After the black
hole is gone, the origin of spherical coordinates appears shifted over
in the diagram.  No spacelike slice can enter the upper diamond
(region $F$) and still be a Cauchy\index{Cauchy surface} surface,
since causal curves that cross the horizon into the black hole will
never reach such a surface. This is the basis of the claim that only
incomplete information is available to observers outside the horizon
after the black hole is gone.

Outside the matter the spacetime is the static Schwarzschild
line\index{line element} element, which can be analytically extended
to a complete spacetime, the diagram of which is given in
figure~\ref{F:CP}\,(e).  This so-called ``eternal'' black hole
spacetime is time-symmetric, with a white hole\index{white hole}
singularity in the past to match the black hole singularity in the
future.  It has two asymptotic spatial regions, connected through a
``throat''\index{throat} at the center of the diagram.

\subsection{Diagrams of static and uniformly moving walls}
\subsubsection{Static wall}
\begin{figure}[tbh]
\centerline{
\psfig{figure=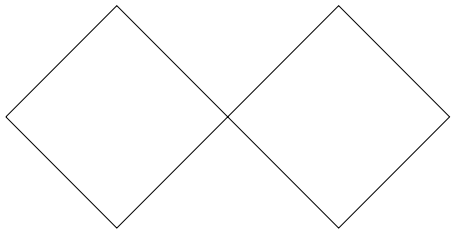,angle=0,height=3cm}}
\caption[Static wall]{\sl Static wall.}
\label{F:statwall}
\end{figure}

As explained in section~\ref{S:statwall} the spacetime of the static
wall is just two complete copies of Minkowski spacetime.  The causal
diagram is figure~\ref{F:statwall}. In strictly relativistic terms
there is no connection at all between the two spacetimes, however the
quasiparticles\index{quasiparticle} can travel
superluminally\index{superluminal} and thus pass from one side of the
wall to the other at a finite value of the time coordinate $t$.  We
indicate this physical connection by depicting the spacetimes as
joined at the wall at spatial infinity.

\subsubsection{Moving wall}
\begin{figure}[p]
\centerline{
\psfig{figure=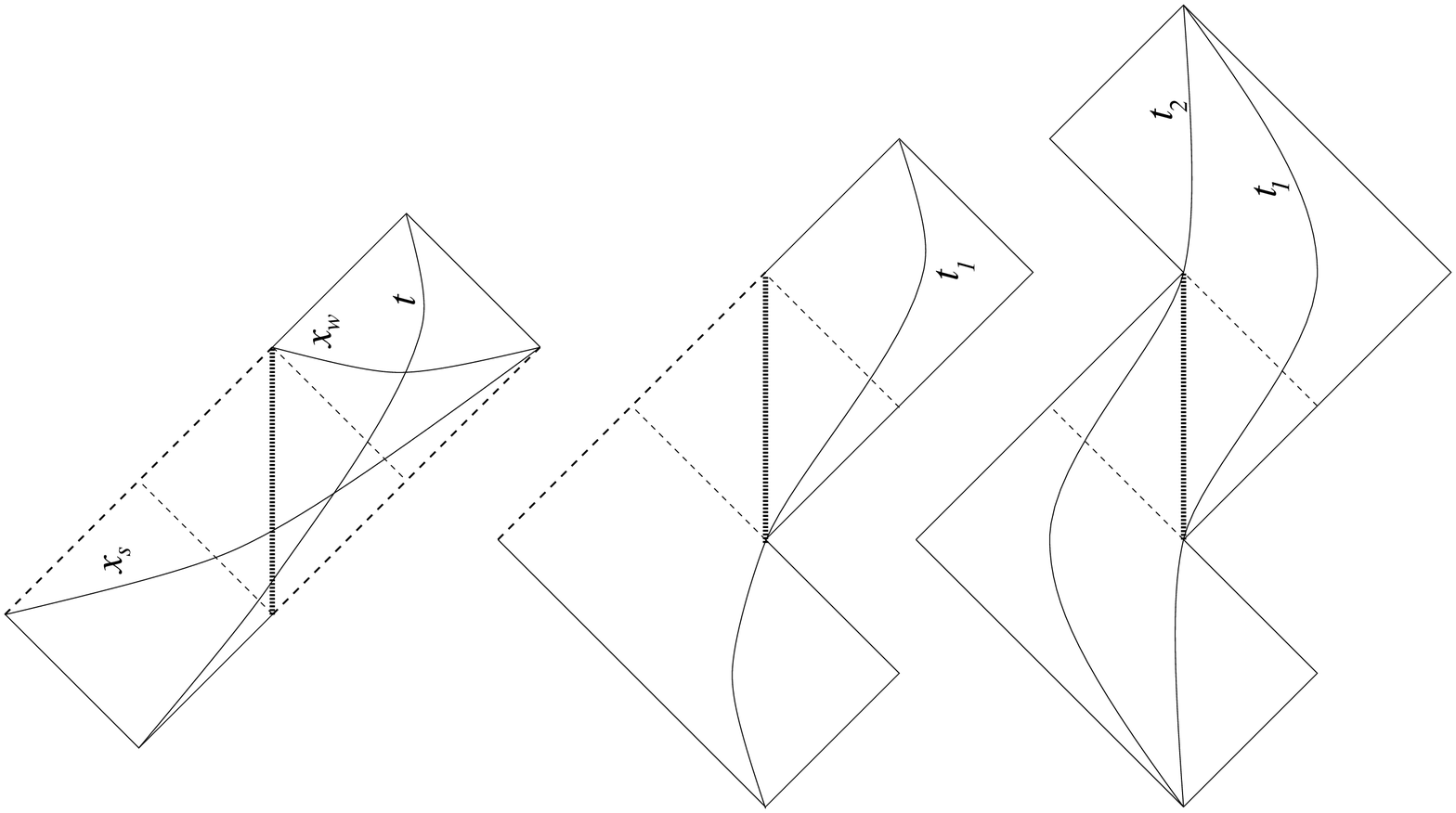,angle=-90,height=18cm}}
\caption[Causal diagrams of moving domain-wall textures.]
{\sl Causal diagrams of moving domain-wall textures.}
\label{F:movewall}
\end{figure}

The causal structure of the spacetime of the moving wall
(\ref{E:movingwall}) is shown in figure~\ref{F:movewall}\,(a).  It
looks like what one would obtain by cutting the diagram for the
eternal black hole (figure~\ref{F:CP}\,(e)) along the white
hole\index{white hole} horizon, and sliding the lower half up so that
the white hole\index{white hole} singularity coincides with the black
hole singularity.  The result of this cut is to leave the spacetime
incomplete along the cut, which corresponds to the pair of long-dashed
lines in figure~\ref{F:movewall}\,(a).  Geometrically it is not
well-defined to extend the spacetime across the singularity, but
physically in the thin film there is continuity in passing through the
core\index{core} of the domain\index{domain wall} wall.  The other
side of the wall thus plays the role of a baby\index{baby universe}
universe.

To clarify the relation between the conformal diagram and the physical
spacetime we include in figure~\ref{F:movewall}\,(a) lines of constant
$t$, $x_s$, and $x_w$.  Note that the incomplete boundaries are at
$t\rar \pm\infty$.  They are the terminus of a null ray that runs
parallel to the black hole horizon backward in time, or the white
hole\index{white hole} horizon forward in time.  Such null rays
asymptotically approach the horizon, blueshifting until
superluminal\index{superluminal} terms in the
dispersion\index{dispersion relation} relation become important, at
which stage a quasiparticle\index{quasiparticle} would cross the
horizon.  Note also that the Newtonian time slices cross the
singularity progressively from left to right.

\subsection{Black hole creation and evaporation}
\subsubsection{Creation}
To construct an analogue of a black hole that forms by collapse we
imagine that the wall velocity $v$ is now a function of Newtonian time
$v(t)$ which is equal to zero before $t_1$ and thereafter smoothly
approaches $v$.  (We do not attempt at this stage to devise a
mechanism for actually accelerating the wall in this manner.)  The
resulting spacetime should look like a portion of the static wall
figure~\ref{F:statwall} below $t_1$ and a portion of the moving wall
figure~\ref{F:movewall}\,(a) above $t_1$.  This yields
figure~\ref{F:movewall}\,(b).  Note that the past incomplete boundary
is now absent because the black hole forms at a finite time.

\subsubsection{Evaporation}
In the case of a real black hole the energy for the
Hawking\index{Hawking effect} radiation comes from the mass of the
hole.  As the hole radiates it loses mass until it evaporates away
completely, unless stabilized by conserved charges it cannot shed.  In
the case of the domain\index{domain wall} wall, the radiation occurs
only when the wall is moving, and it is possible that the
back-reaction to the radiation causes the wall to slow down.  On the
other hand, as discussed in subsection~\ref{SS:hawking}, Pauli
blocking may well terminate the Hawking\index{Hawking effect} process
before the wall comes to rest.  There is presumably another
dissipation mechanism, such as pair-breaking due to the
time-dependence of the moving texture\index{texture}, that eventually
would stop the wall.  In any case, for the purposes of creating a
model of black hole evaporation, we can imagine simply that somehow or
another the wall comes to rest at a time $t_2$.  The resulting
spacetime should then again look like a portion of the static wall
above $t_2$, as shown in figure~\ref{F:movewall}\,(c).

The causal structure for the wall that accelerates and then stops is
similar but not entirely analogous to the spacetime of a black hole
that evaporates.  There is no region analogous to region $F$ of
figure~\ref{F:CP}\,(d), and in fact the spacetime is globally
hyperbolic\index{hyperbolic}.  The analogy is improved if we lean on
the role of the Newtonian simultaneity to define what is accessible
``at a given time''.  Thus, subsequent to $t_2$, the spacetime
consists again of two causally disconnected pieces analogous to $F$
and the black hole interior of figure~\ref{F:CP}\,(d).

\begin{figure}[htp]
\centerline{
\psfig{figure=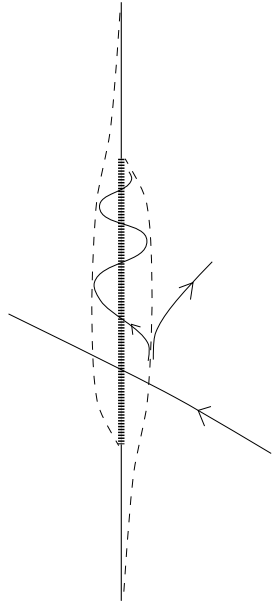,angle=0,height=8cm}}
\caption[Temporary one-way window]{\sl Temporary one-way window.}
\label{F:window}
\end{figure}

To understand better what is going on it is helpful to use also a
non-conformal diagram, in which the domain\index{domain wall} wall
worldline is drawn vertically, figure~\ref{F:window}.  The singularity
appears when the wall starts moving, and disappears when it comes to
rest.  During the motion the singularity is a window to the other side
of the wall.  Figure~\ref{F:window} shows the black and white
hole\index{white hole} horizons as dashed lines, as well as a
quasiparticle\index{quasiparticle} worldline that crosses from right
to left, and a Hawking\index{Hawking pair} pair that is created at the
temporary horizon.
Note that as it approaches the white hole\index{white hole} horizon,
the partner of the Hawking\index{Hawking quantum} quantum is turned
back toward the singularity since it is rightmoving with respect to
the superfluid\index{superfluid} condensate\index{condensate}. It
``rattles'' back and forth between the horizons until the wall stops
moving.

\subsection{Information loss}

It seems clear from the previous discussion that
quasiparticle\index{quasiparticle} information can fall across the
horizon and be lost to the outside world.  The information {\it could}
come back carried by superluminal\index{superluminal} high energy
quasiparticles\index{quasiparticle}, but it {\it need not} and there
is no reason to suppose it does.  The question of information loss by
Hawking\index{Hawking effect} radiation is thornier.  The partners
remain in the ergoregion\index{ergoregion} and fill the negative
energy states.  It seems that roughly half their information would
wind up on the right side of the singularity to be available after the
black hole ``evaporates''.  Still, that leaves the other half that is
lost.

\section{Conclusion}
This is just the beginning of the story.  Clearly a lot remains to be
done to understand the nature of the Hawking\index{Hawking effect}
effect in the setting of the thin-film domain\index{domain wall} wall.
Nevertheless, we hope that this analogue model will prove stimulating to
researchers pondering the nature of Hawking\index{Hawking effect}
radiation and information\index{information loss} loss in
quantum\index{quantum gravity} gravity on the one hand, and the
physics of moving superfluid\index{superfluid} textures\index{texture}
on the other.

\section*{Acknowledgments}
The research described here is based partly on joint work by
G.~Volovik and TJ. We would like to thank G.~Volovik for helpful
discussions, and R.~Parentani for useful comments on a draft of this
article.  Most of this work was done while TK was visiting the Gravity
Theory Group at University of Maryland, for whose hospitality he is
very grateful.  This research was supported in part by the Japan
Society for the Promotion of Science and by the National Science
Foundation under NSF grant PHY-9800967.



\end{document}